# Self-passivated freestanding superconducting oxide film for flexible electronics

Zhuoyue Jia[1#], Chi Sin Tang[2*], Jing Wu[3], Changjian Li[4], Wanting Xu[1], Kairong Wu[1], Difan Zhou[1], Ping Yang[2], Shengwei Zeng[5], Zhigang Zeng[1*], Dengsong Zhang[6], Ariando Ariando[5], Mark B.H. Breese[2,5], Chuanbing Cai[1], Xinmao Yin[1*]

**Affiliations**
[1]Shanghai Key Laboratory of High Temperature Superconductors, Physics Department, Shanghai University, Shanghai 200444, China

[2]The Singapore Synchrotron Light Source (SSLS), National University of Singapore, Singapore 117603

[3]Institute of Materials Research and Engineering, Agency for Science, Technology and Research (A∗STAR), 2 Fusionopolis Way, Singapore, 138634 Singapore

[4]Department of Materials Science and Engineering, Southern University of Science and Technology, Shenzhen, Guangdong, 518055 China

[5]Department of Physics, Faculty of Science, National University of Singapore, Singapore, 117551, Singapore

[6]International Joint Laboratory of Catalytic Chemistry, College of Sciences, Shanghai University, Shanghai 200444

*Correspondence to: slscst@nus.edu.sg (C.S.T.), zgzeng@shu.edu.cn (Z.Z.), yinxinmao@shu.edu.cn (X.Y.)

The integration of high-temperature superconducting $YBa_2Cu_3O_{6+x}$ (YBCO) into flexible electronic devices has the potential to revolutionize the technology industry. The effective preparation of high-quality flexible YBCO films therefore plays a key role in this development. We present a novel approach for transferring water-sensitive YBCO films onto flexible substrates without any buffer layer. Freestanding YBCO film on a polydimethylsiloxane substrate is extracted by etching the $Sr_3Al_2O_6$ sacrificial layer from the $LaAlO_3$ substrate. In addition to the obtained freestanding YBCO thin film having a $T_c$ of 89.1 K, the freestanding YBCO thin films under inward and outward bending conditions have $T_c$ of 89.6 K and 88.9 K, respectively. A comprehensive characterization involving multiple experimental techniques including high-resolution transmission electron microscopy, scanning electron microscopy, Raman and X-ray Absorption Spectroscopy is conducted to investigate the morphology, structural and electronic properties of the YBCO film before and after the extraction process where it shows the preservation of the structural and superconductive properties of the freestanding YBCO virtually in its pristine state. Further investigation reveals the formation of a YBCO passivated layer serves as a protective layer which effectively preserves the inner section of the freestanding YBCO during the etching process. This work plays a key role in actualizing the fabrication of flexible oxide thin films and opens up new possibilities for a diverse range of device applications involving thin-films and low-dimensional materials.

# 1. Introduction

Flexible materials have recently garnered substantial interests due to their remarkable bendability and foldability, leading to widespread utilization in wearable products[1] such as flexible sensors[2] and displays[3]. One area of increasing attention is the study of $YBa_2Cu_3O_{6+x}$ (YBCO) superconducting films for their potential applications[4,5] in flexible electronics. This is due to their underlying physics and technical maturity in fabricating high-quality YBCO thin-film materials. YBCO films possess excellent low-loss properties and are widely used in passive microwave devices such as infrared bolometers[6], filters, and resonators[4]. However, conventional passive microwave devices are generally fabricated by patterning high-temperature superconducting YBCO films epitaxially grown on single crystal substrates such as MgO and $LaAlO_3$ (LAO) [7]. The inherent brittleness of single crystal substrates precludes bending deformation, limiting the incorporation of High-temperature superconductor (HTS) devices into flexible electronic devices and representing a significant obstacle in fulfilling the potential of large-scale utilization and application of HTS-based devices.

Various methods have been employed in recent years to prepare flexible YBCO-based devices such as superconducting wires with low passive loss and radio frequency (RF) properties[8], and microbolometer array devices[6,9]. Among them, the success of wearable micromechanical infrared sensor arrays prepared by flexible YBCO film has been a notable breakthrough[6]. Nevertheless, despite the success of physical release methods in preparing flexible YBCO films for multiple electronic applications, with the discontinuous nature of the naturally layered crystals at the interface, a major prevailing drawback is the structural damage inflicted upon the thin-film YBCO material during the physical release process[8,10]. Thereafter, with the development of new methodologies, the $Sr_3Al_2O_6$ (SAO) has been utilized as a chemical sacrificial layer to prepare freestanding thin-film systems[10,11], where the release method provides greater versatility in the treatment and transfer of thin-film structures while significantly reducing their damage during the process. While the development of this new transfer technique has been a major breakthrough, pressing problems prevail where the mechanical removal of the buffer layer after the etching process may still damage the film to a similar extent as physical release methods[11]. Therefore, it is still challenging to prevent physical damage to the freestanding YBCO film without buffer layers.

Here, we report the implementation of the epitaxial lift-off method using SAO sacrificial layers to produce large-area freestanding YBCO films without using any buffer layers. Using sodium hydroxide (NaOH) solution as an etching agent to remove the SAO sacrificial layer, it effectively inhibits the decomposition of YBCO films where their superconductive properties remained well-preserved. The freestanding YBCO film that we obtained has a $T_c$ of 89.1 K. Interestingly, when subjected to inward and outward bending conditions, the $T_c$ of the freestanding YBCO films is 89.6 K and 88.9 K, respectively. Based on a series of comprehensive characterization techniques comprising X-ray diffraction (XRD), scanning electron microscope (SEM) imaging, transport and X-ray absorption spectroscopic (XAS) characterization, we evaluated the structural and electronic properties and observed the formation of a passivated YBCO layer during the etching process where it further serves as a protective layer which effectively preserves the electronic and superconductive properties of the YBCO film virtually in its pristine state. In comparison with other acid-, water- and alkaline-based etching techniques which will result in the rapid decomposition of the thin-film layer[12], we show that the formation of the YBCO passivated layer can preserve the structural, mechanical and electronic properties of the freestanding YBCO film even after being immersed in the etching solution for four days and beyond. The effective implementation of this technique offers new opportunities in the synthesis and treatment of freestanding membranes beyond chemically stable oxide and perovskite thin-film crystals, to other low-dimensional material systems where applications related to heterostructure electronic architectures and flexible electronics systems[10,13] can be accorded. Moreover, with growing interests related to effects such as those of magic-angle twisted bilayer graphene and other exotic heterostructures attributed to the many-body correlations induced by broken periodic lattice symmetry[14,15], the effective manipulation of such freestanding thin-film structures could also offer new opportunities in the exploration of interfacial effects of such misaligned systems.

## 2. Experimental Procedures

**Epitaxial film Synthesis**

The synthesis of LAO/SAO/YBCO structure has been carried out using pulse-laser deposition (PLD) with the process displayed in Fig. 1(a)[10]. The SAO layer of 55 nm thickness is deposited on the LAO substrate via PLD at laser energy density of 1.6 J/cm$^2$ with a 1 Hz repetition rate at 720 °C under $p_{O_2}$=1.3×10$^{-3}$ Pa. Thereafter, the 280 nm YBCO layer is deposited on the SAO layer at a laser energy density of 4.3 J/cm$^2$ with 3 Hz repetition rate at 850 °C at $p_{O_2}$=20 Pa. The entire synthesis process was carried out in the layer-by-layer mode. Details are provided in Supplementary Material.

**Extracting the freestanding YBCO membrane**

Polydimethylsiloxane (PDMS) substrates have low poor adhesion which makes it easy for transfer the freestanding films to other substrate[16]. A polydimethylsiloxane (PDMS) layer of thickness ~0.5 mm is adhered to the YBCO surface as a support layer. After which, the SAO layer is etched for ~4 days at room temperature with a sodium hydroxide (NaOH) solution of pH 13.0. When the SAO sacrificial layer is completely dissolved, the PDMS/YBCO sample is then separated from the LAO substrate (Fig. 1(a)). The PDMS/YBCO sample has a dimension of ~4×9 mm (Fig. 1(b)), of which, the structural and electronic properties are well-preserved (see discussion later) and flexible (Fig. 1(a)).

The charge transport properties of the YBCO films are then assessed using the physical property measurement system (PPMS) where the temperature-dependent resistance of the YBCO films (the contact electrode positions shown in Fig. S1) before and after lift-off process (Fig. 1(f)) show only a slight variation in $T_c$ from 90 K before transfer to that of 89.1 K for freestanding YBCO – a clear indication that the superconductive property has been preserved with minimal degradation during the lift-off process.

We adhered the YBCO/PDMS structure onto a 0.4 mm thick copper sheet and molded the sample into inward and outward bending states before testing its superconducting properties using PPMS. Surprisingly, the freestanding YBCO films still exhibited superconducting properties under bending conditions (Fig. 1(g)), with a critical temperature of 89.6 K for the inward bending state and 88.9 K for the outward bending state (discuss later).

## 3. Results and discussion
## 3.1 YBCO Structure

The epitaxial nature of YBCO films on SAO/LAO is further confirmed to be of premium quality by cross-section high resolution transmission electron microscopy (HRTEM) studies. The cross-section image (Fig. 1(c)) of the YBCO/SAO/LAO indicate sharp bottom and top interfaces, with a SAO thickness of ~55 nm. Further zoomed in images show that the SAO layer has an epitaxial relationship with the LAO substrate with the *c*-lattice in the out-of-plane direction. The lattice spacing of 0.385 nm corresponds to the (004) lattice spacing belonging to the SAO layer[10], which is close to the LAO (001) *d*-spacing of 0.380 nm. Across the YBCO/SAO interface, the epitaxy relationship is well-maintained. The out-of-plane *d*-spacing of 1.132 nm corresponds to the *c*-lattice parameter of the YBCO lattice[17], and there is no off-tilt grain visible using HRTEM as consistent with the subsequent XRD results.

Figs. 2(a) and (b) show the X-ray diffraction (XRD) patterns of the epitaxial YBCO/SAO structure on LAO (001) substrate. The diffraction patterns belonging to the YBCO and SAO components confirm the single-phase crystallinity epitaxial growth along the *c*-axis of the YBCO/SAO/LAO structure[18]. Wide-angle XRD scans of the freestanding YBCO films presented in Fig. S2(a) show that only a series of YBCO film $c$-axis peaks from (001) to (007) are observed, indicating the complete removal of SAO sacrificial layers and the formation of high-quality freestanding YBCO films after the film extraction process. To further examine the crystal orientation of the of the extracted YBCO films, a fourfold symmetry was also conducted through the in-plane Φ-scan for the (103) YBCO (Fig. 2(c)) before and after the lift-off process[19]. Rocking curve analysis along the (005) YBCO peak further confirms the high quality of the YBCO films (Fig. 2(d)). Collectively, these are clear indications that the crystalline structure of the YBCO thin-film is well-maintained even after the lift-off process[19,20].

## 3.2 Passivated Layer

Having noted the high degree of structural uniformity of the YBCO film along with the preservation of its superconductive property after the lift-off process, it is important to consider how such properties have been preserved through the chemical etching process. NaOH solution of pH 13.0 has an inhibitory effect on the dissolution of YBCO (Supplementary Material). While a NaOH solution of pH 13.0 has sufficient time of ~4 days to decompose the YBCO film during the etching process, we note that no significant changes to the structural and transport properties of the YBCO layer has been observed. Furthermore, the freestanding YBCO layer remains virtually in its pristine condition after 7 days of immersion in the NaOH solution. These results are confirmed using the images as displayed in Fig. 1(b) and Fig. S4. These are clear indications that the YBCO films were not heavily corroded by the NaOH solution. As discussed in greater detail thereafter, the preservation of the pristine condition can be attributed to the formation of a passivated YBCO layer on the surface of the freestanding YBCO film during the etching process to protect the YBCO from any further degradation the NaOH solution.

## 3.3 Formation and contents of the passivated layer

To examine how the formation the YBCO passivated layer takes place, we proceeded to examine the phonon properties, intrinsic electronic structures and the characteristic superconductive phases of the YBCO layer before and after the lift-off process. In this case, Raman spectroscopy serves in as an effective experimental technique to characterize the vibrational modes of the oxygen components and their coupling with the Cu-atoms, in the $CuO_2$ planes and along the Cu-O chain layers[21,22]. Fig. 3(a) shows the Raman spectra of the YBCO films before and after the lift-off process where characteristic peaks at ~336 cm$^{-1}$ are observed. This feature represents the O(2)- O(3) out-of-phase vibrational mode ($B_{1g}$ symmetry) of oxygen of the $CuO_2$ planes (Fig. 3(c))[21]. The peak at ~498 cm$^{-1}$ represents the O(4) vibrational mode ($A_g$) in apical sites of orthorhombic $YBa_2Cu_3O_{6+x}$ (Fig. 3(c))[23] while the vibrational mode O(4) is a superposition of three modes centered near ~474, ~488 and ~500 cm$^{-1}$, respectively, where their relative intensities are typically considered to be dependent on the YBCO oxygen content[24]. In this case, a drop in oxygen concentration will result in a red shift in the O(4)-mode due to a relative intensity increase of the low-energy (~474 cm$^{-1}$) component[23,25]. Besides, the difference in Raman scattering efficiency between the *a*- and *c*-axis grains is attributed to its dependence upon the incident photon angle. The *c*-axis oriented grain fraction (Fig. 3(d)) is calculated to be 0.97 by the relative intensities of O(2,3)-$B_{1g}$ and O(4)-$A_g$ modes (details in Supplementary Material)[26,27]. Thereby indicating the high quality of the YBCO/SAO/LAO structure.

After the lift-off process has been conducted for the YBCO layer, while a significant change takes place to the Raman spectrum, by calculating the *c*-axis oriented grain fraction of the freestanding YBCO film, it only increases marginally to 0.98. This suggests that the YBCO structure remains well-preserved after the lift-off process. Apart from the O(2,3) Raman mode at ~336 cm$^{-1}$ and the red-shifted O(4)-mode (~470 cm$^{-1}$) which were present before the lift-off process, two additional Raman features have also emerged at ~225 and ~590 cm$^{-1}$, respectively – consistent with previous studies where they have been attributed to the vibrations of the Cu(1) and O(1) atoms ($A_g$) along the y-axis (Fig. 3(c)), respectively, near the short Cu-O chains ends[22,23,28]. A slight shoulder feature that emerged at ~250 cm$^{-1}$ is part of the Cu(1) vibrational component along the other axes[22]. With previous studies showing that oxygen-poor tetragonal-phase YBCO have their O(4) Raman mode located at ~470 cm$^{-1}$[22,29], the significant red shift of the O(4) Raman mode in our sample from ~495 to ~470 cm$^{-1}$ upon lift-off could be attributed to the breakage of long-ordered Cu-O chains. Meanwhile, with the absence of characteristic modes from by-product (See Supplementary) such as $BaCuO_2$[30], we can rule out $BaCuO_2$ as the contributing factor to the emergence of the Raman mode at ~590 cm$^{-1}$. The production of other hydrolysis products of YBCO as components of the passivated layer can be further eliminated. With the onset of bond breaking at the YBCO/SAO interface, a large amount of free energy is released[31], which in turn could break the symmetry of the long Cu-O chains[32]. Thus, it is reasonable to claim that the passivated layer is formed by a YBCO layer where a portion of long Cu-O chains have been broken during the etching process.

To confirm that the passivated layer is made up of a YBCO layer with broken Cu-O chains, further characterization process based on X-ray absorption spectroscopy (XAS) is conducted to elucidate the changes to the electronic structures[33] belonging to freestanding YBCO. Figs. 4(a) and (b) compare the polarization-dependent Cu L and O K edges, respectively, of the freestanding YBCO on PDMS and YBCO/LAO at 60° incident angle which allows for the acquisition of electronic structural information along the three crystallographic axes (electric-field, $\vec{\epsilon}/\!/\vec{a}$, $\vec{b}$ and $\vec{c}$). The observed intensities of XAS characteristics in YBCO/LAO for various incident angles, as illustrated in Figs. S5(a) and (b) in Supplementary Material, are found to be highly consistent with those reported for the oxygen-rich $YBa_2Cu_3O_7$ film in a prior investigation[34]. Given that PDMS substrate is a non-conductive polymer material, the signal obtained from the XAS of YBCO/PDMS is relatively weaker. In the Cu $L_3$ edge, the peak intensities along the in-plane $\vec{a}$ and $\vec{b}$ axes do not differ significantly and that most of the contributions can be attributed to orbital contributions along the out-of-plane $c$-axis[34]. Note the significant changes particularly in the form of features A (white line feature at ~931.3 eV attributed to the Cu $3d^9 \rightarrow$ Cu $2p^3d^{10}$ transitions)[34,35], shoulder B (at ~932.3 eV ascribed to the Cu $3d^9L \rightarrow$ Cu $2p^3d^{10}L$ transition in the long Cu-O chains)[34–37] and C (at ~934 eV attributed to the monovalent Cu[34]). The intensities of features A and B for freestanding YBCO/PDMS are lower than that of YBCO/LAO while feature C has become more prominent. These differing Cu $L_3$-edge features between freestanding YBCO/PDMS and YBCO/LAO mirror those observed between oxygen-poor $YBa_2Cu_3O_{6+x}$ and oxygen-rich $YBa_2Cu_3O_7$[34] displayed in Figs. S6(a) and (b). The increase in oxygen (hole) doping concentration leads to the formation of long Cu-O chains particularly along the $b$-axis that leads to significant changes along the out-of-plane direction while changes along the in-plane axes are generally insignificant[34]. This intensity trend of the Cu $L_3$-edge spectra corresponding to the long Cu-O chains in hole-doped $YBa_2Cu_3O_{6+x}$ provides clear evidence that breakage of the long Cu-O chains in a section of the YBCO layer took place during the etching process to form the passivated layer for freestanding YBCO.

The O K-edge spectra of freestanding YBCO/PDMS and YBCO/LAO Fig. 4(b) further confirms the observations made in the Cu L-edge spectra and they also display similar trends as those of oxygen-poor and rich $YBa_2Cu_3O_{6+x}$ corresponding to the presence of long Cu-O chains[34]. This is another clear indication that there are significantly fewer long Cu-O chains in freestanding YBCO/PDMS than in YBCO/LAO – thereby indicating once again the breakage of the long Cu-O chains during the etching process. This YBCO layer with broken Cu-O chains forms the passivated layer which serves as protective layer to prevent any damage to the interior sections of the YBCO from the NaOH solution during the etching process.

## 3.4 The superconducting properties

The freestanding YBCO film shows a marginal decrease in $T_c$ after the lift-off process and an increase in $T_c$ in the bent state. It could be caused by the stress changes along *a-b* plane. With a slightly smaller lattice constant of LAO ($a=b=3.79$ Å) than that of YBCO ($a=3.82$ Å; $b=3.89$ Å)[38,39], there will be a slight compressive strain experienced by the YBCO film on the LAO. Meanwhile, the SAO layer exhibits high lattice flexibility where it adjusts according the lattice parameters of the other epitaxial layers[10]. At the same time, with different thermal expansion of the YBCO, SAO, and LAO layers at $8.6\times10^{-6}$, $9.6\times10^{-6}$ and $9.8\times10^{-6}$ K$^{-1}$, respectively[38,40]. The greater expansion of the YBCO layer leads to a relatively larger compressive stress on the YBCO films when cooled to room temperature. Hence, the YBCO epitaxial layer remains under compressive strain in the YBCO/SAO/LAO system. As YBCO is a ceramic material, its own brittleness leads to cracks during the release of stress. Some microcracks appeared on the freestanding YBCO films during the etching process to release the compressive stress. As such, the anisotropic-uniaxial stress dependence of $T_c$, $dT_c/d\varepsilon$, in the YBCO *a-b* plane along the *a-* and *b-*axis are given by $dT_c/d\varepsilon_a=-230$ K and $dT_c/d\varepsilon_b=220$ K, respectively[38,41,42]. It can therefore be deduced that the marginal decrease in $T_c$ of the freestanding YBCO films can be attributed to the collective effects along the *a-b* plane such that the drop in $T_c$ brought about by the compression along the *a*-axis has been offset by an increase in $T_c$ due to the compressive strain along the *b*-axis. Thus, the relaxation of the lattice compression in the *a-b* plane during the release process has resulted in the marginal decrease in $T_c$[38,41].

Interestingly, unexpected results in the superconducting properties were observed in the YBCO films in both inward and outward bending states. The YBCO films in the bent states still maintain their superconducting properties. By observing the cross-section of the sample in bent states using an optical microscope (Figs. S7(a)), the curvature is positive when bending inward. The curvature of the YBCO film in the inward bending state was measured to be 0.052 mm$^{-1}$ by optical microscope. During the inward bending state, an increase in $T_c$ from 89.1 K to 89.6 K was even observed, and the superconducting transition width decreased from 6.1 K to 4.6 K, bringing it closer to the superconducting properties of the YBCO film on the SAO/LAO structure. The improvement in the $T_c$ of YBCO film is attributed to the compressive stress on the *a-b* plane when the YBCO films in an inward bending state. The results demonstrate that YBCO film in the YBCO/SAO/LAO structure is subjected to compressive stress, and the partial release of this stress during the lift-off process leads to a decrease in the $T_c$. The curvature of the YBCO film (Figs. S7(b)) in the outward bending state was measured to be -0.123 mm$^{-1}$. During the outward bending state, a slight decrease of $T_c$ from 89.1 K to 88.9 K was observed, and the superconducting transition width decrease from 6.1 K to 4.9 K. Since there have been no reports on the effect of tensile strain of YBCO films on $T_c$ in previous studies[41,42], the reduction in $T_c$ of the YBCO film is attributed to the release of residual compressive stress in the freestanding YBCO film in an outward bending state.

The same trend in changes to the superconducting property was also observed in other sample groups. Such as in Fig S8, with a 1 K increase in the $T_c$ of the YBCO film under inward bending conditions and a 0.2 K decrease under outward bending state. Due to the larger curvature from 0.052 mm$^{-1}$ to 0.072 mm$^{-1}$ of the inward bending state, the YBCO film in this sample group experienced a larger compressive strain, resulting in an increased change $T_c$ by 0.5 K to 1 K, consistent with previous studies[41]. The results above demonstrate that the YBCO films in bent states still possess excellent superconducting performance, representing a significant step forward in the integration of YBCO high-temperature superconducting films with flexible electronic devices.

# 4. Conclusion

In summary, we present a novel approach to fabricate freestanding YBCO films without the use of protective buffer layers. Experimental characterization studies show the formation of a passivated layer which protects and effectively preserves the structural and superconductive properties of the YBCO film from the caustic effects of the etching solution. We further observe that freestanding YBCO films still exhibit a $T_c$ of around 90 K under bending conditions. Our preparation methodology highlights the effective synthesis and treatment of freestanding membranes that transcends chemically stable oxide and perovskite thin-films crystals into other thin-film and low-dimensional materials systems their integration into applications related to heterostructure electronic architectures and flexible electronics systems[10,13].

**Acknowledgements:**

This work was supported in part by the Strategic Priority Research Program of the Chinese Academy of Sciences, Grant No. XDB25000000, National Natural Science Foundation (52172271), Shanghai Science and Technology Innovation Program (22511100200). C. S. T. acknowledges the support from the NUS Emerging Scientist Fellowship. J. W. acknowledges the Advanced Manufacturing and Engineering Young Individual Research Grant (AME YIRG Grant No.: A2084c170) and the SERC Central Research Fund (CRF). The authors would like to acknowledge the Singapore Synchrotron Light Source for providing the facility necessary for conducting the research. The Laboratory is a National Research Infrastructure under the National Research Foundation, Singapore. Any opinions, findings and conclusions or recommendations expressed in this material are those of the author(s) and do not reflect the views of National Research Foundation, Singapore.

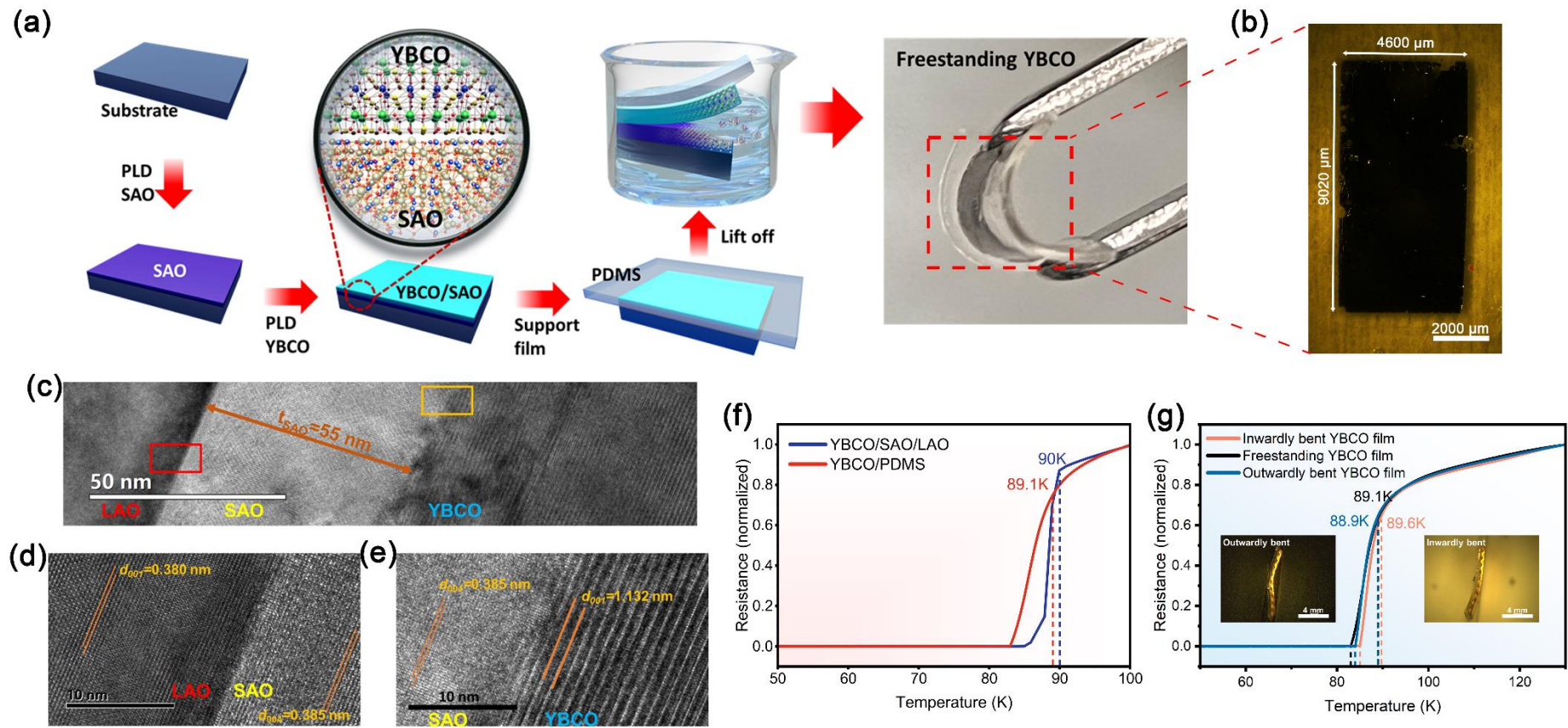

**FIG. 1.** (a) Process schematic for YBCO/SAO/LAO structure growth and the lift-off of freestanding YBCO film. HRTEM images of YBCO/SAO//LAO structure. (c) An overview of the YBCO/SAO//LAO cross-sectional image. (b) Optical microscope photograph of YBCO freestanding film on PDMS. The zoomed image of the bottom LAO/SAO (d, red rectangle enclosed region in c) and top YBCO/SAO interface (e, orange rectangle enclosed region in c), respectively. (f) Comparison of the temperature-dependent resistance data between YBCO/SAO/LAO

structure and freestanding YBCO films, normalized relative to their resistance at 100 K.  (g) Comparison of the temperature-dependent resistance data among freestanding, bend inward and outward YBCO films, normalized relative to their resistance at 130 K.

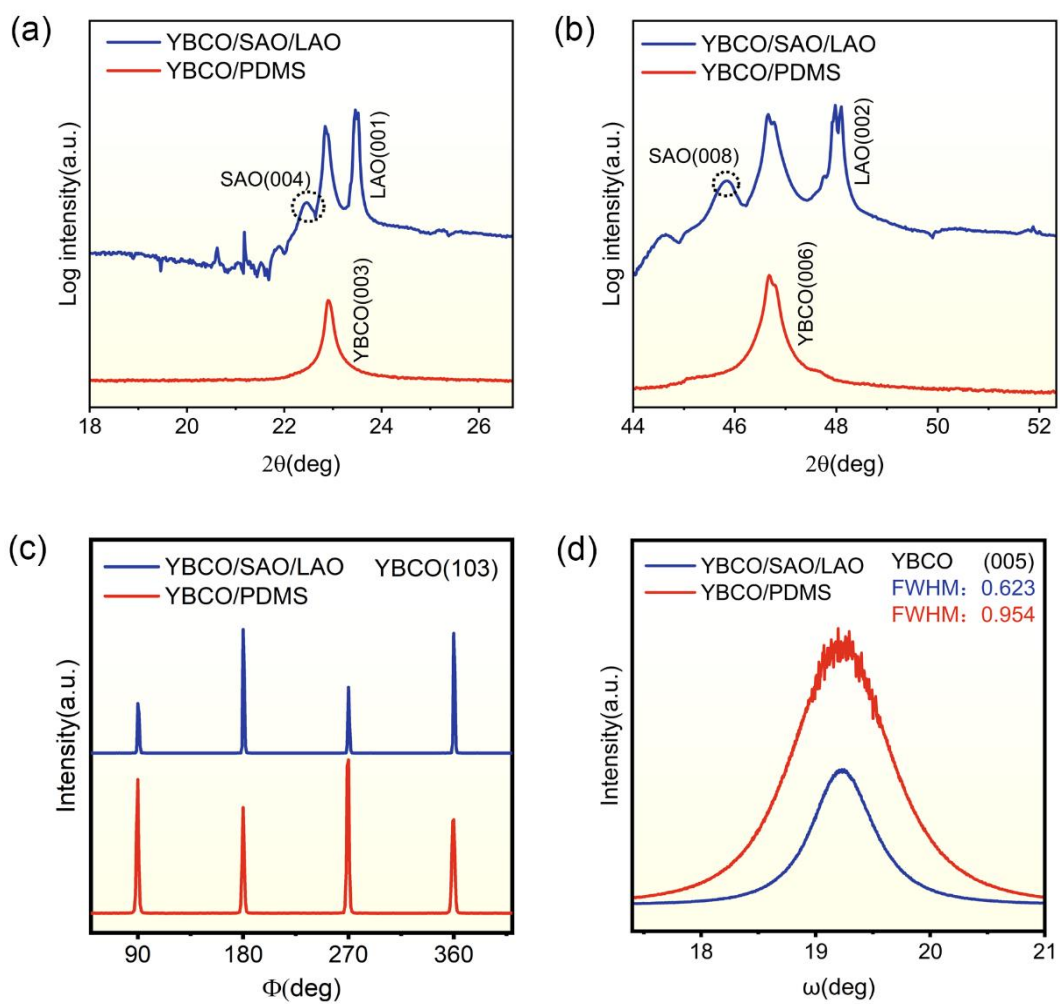

**FIG. 2.** (a) and (b) The characteristic peaks XRD pattern of epitaxial YBCO/SAO structure on LAO (001) substrate and freestanding YBCO film on PDMS. (c) XRD Φ-scan for YBCO (103) peak. (d) XRD ω-scan (rocking curve) for YBCO (005) peak.

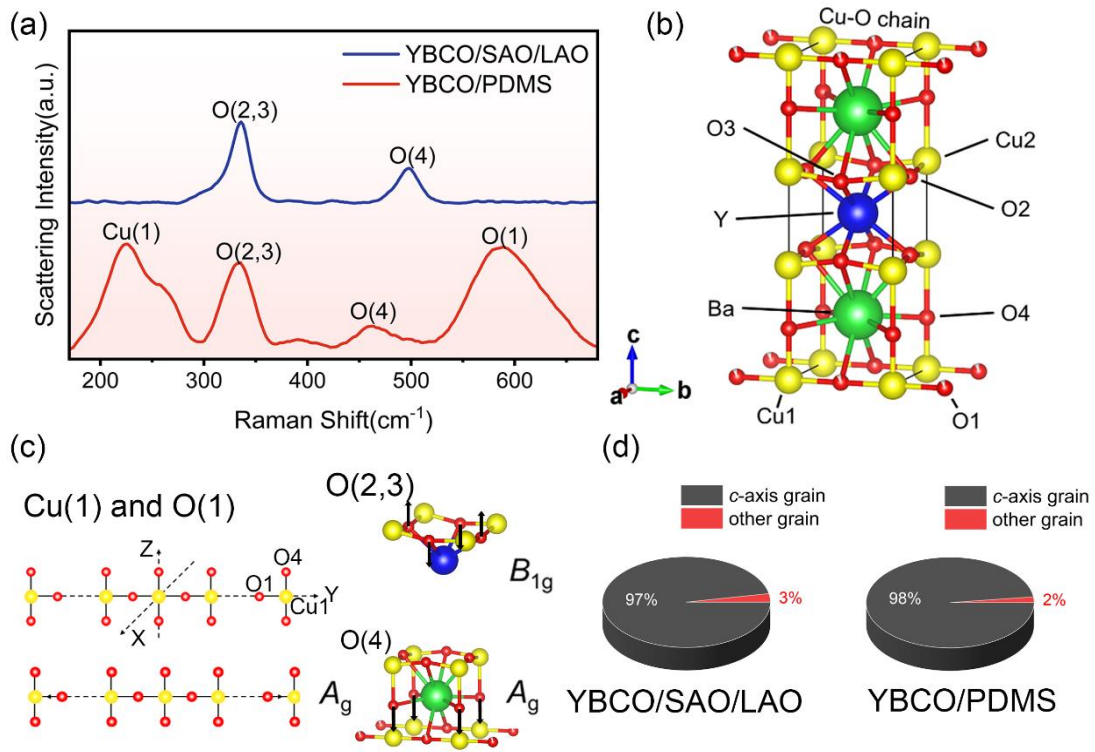

**FIG. 3.** (a) Raman spectra of YBCO film before and after lift-off process. (b) The structure of YBCO in orthorhombic phase. (c) The Raman vibrational modes of Cu(1), O(1), O(2,3) and O(4) in YBCO. (d) The $c$-axis oriented grain fraction calculated by Raman spectra.

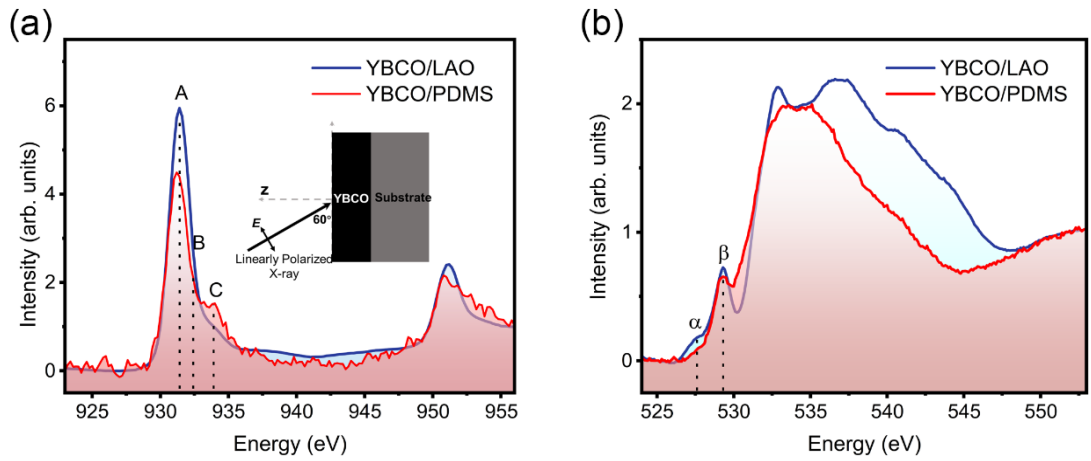

**FIG. 4.** The x-ray absorption spectra of YBCO on LAO and PDMS at the (a) Cu $L_3$ edge and (b) Oxygen K edge for 60 deg.